%% file: main.tex
\documentclass[12pt, utf8]{article}
\usepackage[body={17cm, 23cm, centered}]{geometry}
\usepackage[english]{babel}

\usepackage[utf8]{inputenc}

\usepackage{amsmath,amssymb,amsthm,amscd,cite,comment,amsfonts,indentfirst,color,setspace,bbold,arydshln}
\usepackage{mathtext,marvosym,textcomp}
\usepackage{lmodern}

\usepackage[makeroom]{cancel}
\usepackage{mathtools}
\usepackage{braket}
\usepackage{cancel}
\usepackage{float}
\usepackage{bbm}

\usepackage[debug,pageanchor=false]{hyperref}
\hypersetup{colorlinks=true,linktocpage,breaklinks,
	urlcolor=blue,
	linkcolor=blue,
	citecolor=blue
}

\usepackage[vcentermath]{youngtab}

\usepackage[mathmode,centertableaux]{ytableau}
\usepackage{tikz}
\usetikzlibrary{arrows}
\usetikzlibrary{shapes.misc}
\usepackage{tikz-cd}

\usepackage[ normalem]{ulem}
\usepackage{tocloft}

\numberwithin{equation}{section}

\selectfont

\input{Defs.tex}

\begin{document}
\renewcommand{\contentsname}{}
\renewcommand{\refname}{\begin{center}References\end{center}}
\renewcommand{\abstractname}{\begin{center}\footnotesize{\bf Abstract}\end{center}} 

\begin{titlepage}
\ph{preprint}

\vfill

\begin{center}
   \baselineskip=16pt
   {\large \bf Poly-vector deformations of heterotic supergravity solutions
   }
   \vskip 2cm
    Kirill Gubarev$^{a,b,c}$\footnote{\tt kirill.gubarev@phystech.edu }, Konstantin Sovit$^c$\footnote{\tt sovit.km20@physics.msu.ru}
       \vskip .6cm
             \begin{small}
                          {\it
                          $^a$Institute for Information Transmission Problems, 127051, Moscow, Russia\\
                          $^b$Moscow Institute of Physics and Technology, Laboratory of High Energy Physics\\
                          Institutskii per., 9, 141702, Dolgoprudny, Russia\\ 
                          $^c$Institute of Theoretical and Mathematical Physics, Lomonosov Moscow State University, \\Moscow 119991, Russia    } \\ 
\end{small}
\end{center}

\vfill 
\begin{center} 
\textbf{Abstract}
\end{center} 
\begin{quote}
We construct bi- and uni-vector deformations of 10d heterotic supergravity solutions with the gauged double
field theory approach. We construct a generalization of the ``open/closed'' map for this case and consider some examples of the deformed solutions, particularly for the F1 string solution. 
\end{quote}

\vfill
\setcounter{footnote}{0}
\end{titlepage}

\tableofcontents

\setcounter{page}{2}

\section{Introduction}

Supergravity solution generating techniques provide powerful methods for investigating string and M-theory. Among the most well-known today are T-, S- and U-dualities \cite{PhysRevLett.58.1597,MONTONEN1977117,Cremmer:1997ct,Cremmer:1998px,HULL1995109}, Poisson–Lie (non-abelian) T-duality and T-plurality \cite{Klimcik:1995jn,CtiradKlimcík_2002,KLIMCIK1995455,VonUnge:2002xjf}, Nambu–Lie (non-abelian) U-duality \cite{10.1093/ptep/ptz172,Malek:2019xrf,Sakatani:2020wah,Malek:2020hpo,PhysRevD.104.046015,Musaev:2020bwm}, abelian and non-abelian fermionic T-duality \cite{Berkovits:2008ic,Bakhmatov:2011ab,NIKOLIC2017105,Astrakhantsev:2021rhj,Astrakhantsev:2022mfs,Osten:2016dvf}, Yang–Baxter deformations \cite{Bena:2003wd,Klimcik:2002zj,Klimcik:2008eq} and generalized Yang–Baxter deformations, also known as poly-vector deformations \cite{Ashmore:2018npi,Lunin:2005jy,Bakhmatov:2019dow,Bakhmatov:2020kul,Gubarev:2020ydf}.

Poly-vector deformations occupy a special place in the list above and have attracted significant attention in recent years. Such deformations were first studied for the NS-NS sector of 10D supergravity and are known as bi-vector deformations \cite{Bakhmatov:2017joy,Bakhmatov:2018apn,Bakhmatov:2018bvp}. They are given by a generalization of the open/closed string map \cite{Seiberg:1999vs}
\begin{equation}
\label{eq:deformation0}
    g+b = \big((G+B)^{-1} + \beta\big)^{-1}, \quad e^{-4 \phi} \det g = e^{-4 \Phi} \det G,
\end{equation}
where g, b and $\phi$ are the deformed metric, Kalb–Ramond field, and dilaton, respectively; G, B and $\Phi$ are the initial ones; and $\beta$ is the deformation bi-vector. It was found that they generate new backgrounds from supergravity solutions if the bi-vector is taken to be in the bi-Killing ansatz, $\beta^{mn} = r^{ij} k_{i}{}^{m} k_{j}{}^{n}$, and the constant antisymmetric $r$-matrix satisfies \cite{Bakhmatov:2018bvp}
\begin{equation}\label{10dcond}
    \begin{cases}
         f_{j_1 j_2}\,^{[i_1} r^{i_2|j_1|} r^{i_3]j_2} = 0, \qquad &\text{(CYBE)},\\
         f_{i_1 i_2}\,^{j} r^{i_1 i_2} = 0, \qquad &\text{(unimodularity)}.
    \end{cases}
\end{equation}
where $f_{i_1 i_2}{}^{i_3}$ are the structure constants of the Killing vectors algebra, $[k_{i_1},k_{i_2}] = f_{i_1 i_2}{}^{j} k_{j}$.

Moreover, bi-vector deformations are related to the preservation of integrability \cite{Klimcik:2002zj,Klimcik:2008eq,Delduc:2013qra} and kappa-symmetry \cite{Arutyunov:2015qva,Arutyunov:2015mqj} of the 2D sigma-models on the corresponding backgrounds. It is worth mentioning that even without the unimodularity condition in (\ref{10dcond}), kappa-symmetry is preserved, but the deformed background becomes a solution of generalized supergravity \cite{Bakhmatov:2017joy,Arutyunov:2015mqj,Wulff:2016tju,Sakatani:2016fvh} with $I^m = \nabla_{k}\beta^{km}=f_{i_1 i_2}\,^{j} r^{i_1 i_2}k_{j}{}^{m} \neq 0$. From the perspective of the open/closed string map, these deformations are interpreted as linking commutative and non-commutative theories with non-commutative parameter $\beta$, that is well understood only in the case of constant $\beta$ \cite{Arfaei:1997hb,Sheikh-Jabbari:1997qke} (for a recent progress in this direction see \cite{Barakin:2025jwp}, also methods of \cite{Osten:2019ayq} can be easily generalized to non-constant beta).

Double field theory (DFT) has proven to be crucial for understanding the highly non-linear rule (\ref{eq:deformation0}) for bi-vector deformations \cite{Geissbuhler:2013uka,Aldazabal:2013sca,Bakhmatov:2018bvp}. It provides an O(D,D)-covariant description of IIA/IIB supergravity, and in its framework the deformation becomes simple: it is just a local O(D,D) rotation of the generalized metric,
\begin{equation}
    \mH' = 
        \begin{bmatrix}
            G - B G^{-1} B & G^{-1} B \\
            - B G^{-1} & G^{-1}
        \end{bmatrix} =
        O^T_\b
        \underbrace{
        \begin{bmatrix}
            g - b g^{-1} b & g^{-1}b \\
            - b g^{-1} & g^{-1}
        \end{bmatrix}
        }_{\mH} O_\b
        , \quad
    O_\b = 
        \begin{bmatrix}
            \mathbb{1} & \b \\
            \mathbb{0} & \mathbb{1}
        \end{bmatrix}.
\end{equation}
and symmetry of invariant dilaton
\begin{equation}
    d' = \Phi - \frac{1}{4} \det G = \phi - \frac{1}{4} \det g = d.
\end{equation}

In recent years, this approach to bi-vector deformations from the perspective of extended field theories has been widely extended. As a result, using exceptional field theories (ExFT), tri- and six-vector deformations of 11D supergravity solutions \cite{Bakhmatov:2019dow,Bakhmatov:2020kul,Gubarev:2020ydf,Barakin:2024rnz}, as well as quadri-vector deformations of 10D supergravity solutions \cite{Gubarev:2024tks}, have been constructed. Furthermore, using GL(D+1) general relativity, deformations of solutions of Einstein–Maxwell–dilaton theories have been constructed \cite{Gubarev:2025hvr} (which can be easily extended to GL(D+n)). In all these cases the deformation parameters use a poly-Killing ansatz and, except for the GL case, must satisfy the generalized Yang–Baxter equation and unimodularity conditions to generate solutions.

These poly-vector deformations, like the bi-vector ones, are assumed to be related to the preservation of integrability \cite{Gubarev:2023jtp} and kappa-symmetry \cite{Bakhmatov:2022lin,Bakhmatov:2022rjn,Gubarev:2023xaq} of higher-dimensional sigma models describing M2- and M5-branes. Moreover, they are a useful tool for studying (S)QFT via the holographic correspondence. By this, we mean that deformations on the gravity side of the correspondence are dual to the addition of (ir)relevant or marginal operators, changing the microscopic behavior by introducing non-commutativity, or by considering a phase with a non-vanishing VEV of an operator \cite{Leigh:1995ep,Argyres:1995jj,Lunin:2005jy,vanTongeren:2015uha,Imeroni:2008cr}. This allows the construction of more physically interesting theories, possessing fewer symmetries, from supersymmetric, conformal ones, and also enables calculations in them using the known gravity side.

As mentioned above, for heterotic supergravity, only bi-vector deformations of the NS-NS sector are known. However, heterotic supergravity has a formulation in terms of an extended field theory called gauged double field theory (GDFT) \cite{Grana:2012rr,Aldazabal:2013sca}. Constructing the deformation in this formalism allows to go beyond bi-vector deformations of the NS-NS sector, descending from the ordinary DFT approach. The main goal of this paper is to show that the GDFT approach allows for the construction of a composition of uni- and bi-vector deformations for heterotic solutions, turning on non-trivial vector and Kalb–Ramond fields.

This paper is structured as follows. In Section~\ref{sec:GDFT}, we set up our notation and review the GDFT formulation of heterotic supergravity, its equations of motion, and flux formulation. In Section~\ref{sec:deformation}, we construct uni- and bi-vector deformations of heterotic supergravity solutions as local O(D,D+n) rotations, derive the deformation rules, generalizing the open/closed string map, and discuss constraints on the deformation parameters. In Section~\ref{sec:examples}, we construct examples of deformed backgrounds for flat space and the F1 string. Finally, in Section~\ref{sec:conclusion}, we discuss the obtained results, outline open questions, problems, and directions for further investigation. Calculations based on computer algebra programs Wolfram Mathematica and Cadabra \cite{Peeters:2006kp,Peeters:2018dyg} can be found in Cadabra and Mathematica files on the GitHub repository \cite{Gubarev:2025heterotic}.

\section{Embedding of heterotic supergravity in GDFT}\label{sec:GDFT}

It is known that type II string and supergravity theories compactified on a D-torus enjoy O(D,D;$\mathbb{Z}$) and O(D,D;$\mathbb{R}$) symmetries correspondingly. This allows the description of their NS-NS degrees of freedom in terms of O(D,D) multiplets of DFT \cite{Geissbuhler:2013uka,Aldazabal:2013sca}, where the symmetry is realized geometrically and becomes local. However, the heterotic string has a bigger hidden symmetry group that is O(D,D+n) \cite{Grana:2012rr,Aldazabal:2013sca}. Here n = 496 for both SO(32) and E$_{8}\times$E$_{8}$ heterotic theories. Let us now review the embedding of the heterotic supergravity in the GDFT.

We consider a dimensionally reduced O(D,D+n) theory living in the space with coordinates $X^M =\left( x^{m},  \tilde{x}_{m},  y^{\alpha} \right)$, where $\left( x^{m},  \tilde{x}_{m} \right) = \mathbb{X}$ are the external doubled space coordinates and $y^{\alpha} = \mathbb{Y}$ are the internal coordinates. After the Scherk-Schwarz compactification of the $\mathbb{Y}$-space, we get the action \cite{Aldazabal:2013sca}
\begin{equation} \label{GDFT_flux_action}
\begin{aligned}
S_{\text{GDFT}} = &\ v \int d \mathbb{X} e^{-2  {d}} && \left[\frac{1}{4}\left( {F}_{ABC}+f_{ABC}\right)\left( {F}_{A_1 B_1 C_1}+f_{A_1 B_1 C_1}\right) S^{A A_1} \eta^{B B_1} \eta^{C C_1}  \right. \\
 & && -\frac{1}{12}\left( {F}_{ABC}+f_{ABC}\right)\left(  {F}_{A_1 B_1 C_1}+f_{A_1 B_1 C_1}\right) S^{A A_1} S^{B B_1} S^{C C_1} \\
 & && -\frac{1}{6} \left( {F}_{ABC}+f_{ABC}\right)\left(  {F}_{A_1 B_1 C_1}+f_{A_1 B_1 C_1}\right) \eta^{A A_1} \eta^{B B_1} \eta^{C C_1} \\
  & && \left.+ \left(  S^{A B} - \eta^{A B}  \right)  {F}_A  {F}_B\right] \, . 
\end{aligned}
 \end{equation}
In the above action nothing depends on $y^{\alpha}$, and $v$ is a constant appearing from the integration over the reduced coordinates $\mathbb{Y}$ (see \cite{Aldazabal:2013sca} for details), $d$ is O(D,D+n) invariant dilaton, $S^{A B}$ is inverse to flat generalized metric $S_{A B}$, and $\eta^{A B}$ is inverse to O(D,D+n) invariant metric $\eta_{A B}$ :
\begin{equation}
       S_{A B} =  \left(
   \begin{array}{lll} 
  s_{a b} & 0 & 0\\
  0 & s^{a b} & 0 \\
  0 & 0 & \delta_{\bar{\alpha} \bar{\beta}}
   \end{array}
   \right) \, , \quad 
   \eta_{A B} =
   \left(
   \begin{array}{lll} 
   0 & \delta_a{}^b & 0 \\
   \delta^{a}{}_{b} & 0 & 0 \\
   0 & 0 & \delta_{\bar{\alpha} \bar{\beta}}
   \end{array}
   \right) \, ,
\end{equation}
where $s_{a b} = $ diag$(-+...+)$ is the D-dimensional flat Minkowski metric. Flat indices are transformed to the curved ones by contraction with generalized vielbein $E_{A}{}^{M}\in$  O(D,D+n) and are lowered and raised by O(D,D+n) invariant metric $\eta$. Curved generalized metric is defined as $ {\mathcal{H}}_{M N} =  {E}^A{}_M  {E}^B{}_N S_{A B}$ and fluxes with the flat indices are related to the curved ones as $f_{A B C} = E_{A}{}^{M} E_{B}{}^{N} E_{C}{}^{K} f_{M N K}$, $F_{A B C} = E_{A}{}^{M} E_{B}{}^{N} E_{C}{}^{K} F_{M N K}$, where
\begin{equation}
     {F}_{M N K}=3  {\Omega}_{[A B C]}  {E}^A{}_M {E}^B{}_N {E}^C{}_K, \quad  {\Omega}_{A B C}= {E}_{A}{}^R \partial_R  {E}_{B}{}^S  {E}_{C S} \, , 
\end{equation}
\begin{equation}
    {F}_M = {\Omega}^B{}_{B A} E^A{}_M + 2\partial_{M}{d} \, .
\end{equation}
We assume that the section constraint  $\partial_N  {V} \partial^N  {W}=0$ holds for all gauge parameters, fields and their products $ {W}$ and $ {V}$ depending on the external coordinates. The constant  gaugings  $f_{M N K} = f_{[MNK]}$ are  such that the quadratic constraint $f_{S[M N} f_{K L]}{}{}^S=0$ holds and for the Lorentz invariance of the theory the only nonzero components are $f_{\alpha \beta \gamma}$, which are structure constants of SO(32) or E$_8 \times$E$_8$. Letters from the beginning of the Latin alphabet correspond to the tangent space, a vector in the tangent space  having components $ v^A = \left( v^a \, , v_a \, , v^{\bar{\alpha}} \right)$.  Equations of motion can be obtained by varying with respect to the generalized dilaton $d$ and to the generalized vielbein $E_{A}{}^{M}$, taking into account the fact that it is an O(D,D+n) matrix, meaning that, $ \delta E_{A B} \equiv E_{A M} \delta E_B{}^M = - E_B{}^M \delta E_{A M} $ is antisymmetric:

 \begin{equation}    \label{GDFT_eom}
     \begin{aligned}
\delta d: \quad   &  2 \partial_{A}{F_{B}} \left( S^{A B} - \eta^{A B} \right) + \dfrac{1}{4} Q_{A B C} Q_{D E F} \eta^{A D} \eta^{B E} S^{C F} - \dfrac{1}{12} Q_{A B C} Q_{D E F} S^{A D} S^{B E} S^{C F} \\
& -  \dfrac{1}{6} Q_{A B C} Q_{D E F} \eta^{A D} \eta^{B E} \eta^{C F} - F_{A} F_{B} \left( S^{A B} - \eta^{A B}     \right)  = 0 \, ,  \\ 
\delta E_{A B}: \, &  - \partial_{F}{Q_{C D E}} \eta^{A C} \eta^{B D} S^{E F} 
 + Q_{C D E}  F_{F} \eta^{A C} \eta^{B D} S^{E F}
 - \partial_{F}{Q_{C D E}} \eta^{B C} \eta^{D F} S^{A E} \\
&  + \partial_{F}{Q_{C D E}} \eta^{A C} \eta^{D F} S^{B E} 
 - Q_{C D E} F_{F} \eta^{A C} \eta^{D F} S^{B E}
 + \frac{1}{2} Q_{A_1 B_1 C} Q_{D E F} \eta^{B A_1} \eta^{B_1 D} \eta^{C E} S^{A F} \\
&  - \dfrac{1}{2} Q_{A_1 B_1 C} Q_{D E F} \eta^{A A_1} \eta^{B_1 D} \eta^{C E} S^{B F} 
  + \partial_{F}{Q_{C D E}} S^{A C} S^{B D} S^{E F}
- Q_{C D E} F_{F} S^{A C} S^{B D} S^{E F} \\
& - \dfrac{1}{2} Q_{A_1 B_1 C} Q_{D E F} \eta^{B A_1} S^{A D} S^{B_1 E} S^{C F} 
+ \dfrac{1}{2} Q_{A_1 B_1 C} Q_{D E F} \eta^{A A_1} S^{B D} S^{B_1 E} S^{C F} \\
& - 2 Q_{C D E} F_{F} \eta^{A C} \eta^{B D} \eta^{E F} 
 - 2 \partial_{D}{F_{C}} \eta^{A D} S^{B C} + 2 \partial_{D}{F_{C}} \eta^{B D} S^{A C} + 2 \partial_{D}{F_{C}} \eta^{A D} \eta^{B C}  \\
 & - 2 \partial_{D}{F_{C}} \eta^{A C} \eta^{B D}  + Q_{C D E} F_{F} \eta^{B C} \eta^{D F} S^{A E}    + 2 \partial_{F}{Q_{C D E}} \eta^{A C} \eta^{B D} \eta^{E F} = 0 \, ,
     \end{aligned}
 \end{equation} 
 where $Q_{A B C} = F_{A B C} + f_{A B C} \, $.
 The generalized vielbein can be Lorentz transformed to the form corresponding to the heterotic supergravity parameterization\footnote{Other parameterizations of GDFT can be used to describe non-geometric backgrounds.}
 
\begin{equation} \label{vielbein_generalized}
     {E}_{A}{}^{M} = 
    \left(\begin{array}{ccc} {e}_{a}{ }^{m} & 0 &  0 \\ 
     - {e}_{a}{ }^{l}  {c}_{l m} &  {e}^{a}{ }_{m} &  {A}_{m \bar{\alpha}} \\ 
    - {e}_{a}{ }^{l}  {A}_{l}{}^{\beta } &  0  &   \delta_{\bar{\alpha}}{}^{\beta} \end{array}\right) \, , \quad
        {E}^{A}{}_{M} = 
       \left(\begin{array}{ccc}
      {e}^{a}{ }_{m} &  - {e}_{a}{ }^{l}  {c}_{l m} &  {A}_{m}{}^{\bar{\beta}} \\ 
  0 &   {e}_{a}{ }^{m} & 0 \\ 
    0 &   - {e}_{a}{ }^{l}  {A}_{ l \alpha} &   \delta_{\alpha}{}^{\bar{\beta}} \end{array}\right) \, ,
\end{equation}
associated with the  generalized metric ${\mathcal{H}}_{M N} =  {E}^A{}_M  {E}^B{}_N S_{A B} \, $, 
\begin{equation}
     {\mathcal{H}}_{M N} =
    \left(\begin{array}{ccc}
    {g}_{m n}+ {A}_m{}^\gamma   {A}_{n \gamma }+ {c}_{l m}  {g}^{l s}  {c}_{s n}  &- {g}^{n l}  {c}_{l m} &  {A}_m{}_\beta + {A}_{l \beta }  {g}^{l s}  {c}_{s m} \\
 - {g}^{m l}  {c}_{l n}    &  {g}^{m n} & - {g}^{m l}  {A}_{ l \beta}  \\  {A}_{n \alpha} + {A}_{ l \alpha}  {g}^{l s}  {c}_{s n} & -  {g}^{n l}  {A}_{ l \alpha} & \delta_{\alpha \beta} + {A}_{ l \alpha}  {g}^{l s}  {A}_{ s \beta}
    \end{array}\right) \, ,
\end{equation}
where $ {e}_a{}^n $ is the D-dimensional vielbein associated with $s_{ab}$ and $ {c}_{m n} = {b}_{m n} + \dfrac{1}{2}  {A}_{ m \gamma}  {A}_n{}^\gamma  \, $, and indices $\alpha, \beta$ are lowered and raised by $\delta_{\alpha\beta}$ and $\delta^{\alpha\beta}$. Finally, the invariant dilaton in terms of supergravity fields takes the following form
\begin{equation}\label{invariant_dilaton}
    e^{-2d} = \sqrt{g} e^{-2\phi}.
\end{equation}
The heterotic supergravity action can be obtained from (\ref{GDFT_flux_action}) by substituting the fluxes $F_{A B C}$ using the parametrization of the generalized vielbein (\ref{vielbein_generalized}) and the invariant dilaton (\ref{invariant_dilaton}) and solving the section constraint  by only leaving the dependence on coordinates $ x^m$ ($\frac{\partial}{\partial \tilde{x}_n} = 0$):
\begin{equation}
    \begin{aligned}
     F_{a b}{}^{c} =&\  f_{a b}{}^{c} \, , \\
     F_{a b c} =&\  - e_{a}{}^{m} e_{b}{}^{n} e_{c}{}^{k} H_{m n k} \, , \\
     F_{a b \alpha} =&\  - e_{a}{}^{m} e_{b}{}^{n} F_{m n \alpha} \, , \\
     F_{a \alpha \beta} =&\  - f_{\alpha \beta \gamma} A_{m}{}^{\gamma} e_{a}{}^{m} \, , \\
     F_{\alpha \beta \gamma} =&\  f_{\alpha \beta \gamma} \, , \\
     F_{a} =&\  f_{a b}{}^{b} + 2 \partial_{m}{\phi} e_{a}{}^{m} \, , 
    \end{aligned}
\end{equation}
where 
\begin{equation}
    \begin{aligned}
    f_{a b}{}^{c} =&\ -2  e_{a}{}^{m} e_{b}{}^{n} \partial_{[m} e_{n]}{}^{c} \, ,    \\
    H_{m n k} =&\ 3\left(\partial_{[m} B_{n k]}-\delta_{\alpha \beta} A_{[m}^\alpha \partial_n A_{k]}^\beta\right)+ f_{\alpha \beta \gamma} A_{[m}{ }^\alpha A_n^\beta A_{k]}^\gamma \, , \\
    F_{m n \alpha} =&\ 2 \partial_{[m} A_{n] \alpha} - f_{\alpha \beta \gamma } A_{[m}^\beta A_{n]}^\gamma \, . 
    \end{aligned}
\end{equation}
After substituting these into the action (\ref{GDFT_flux_action}) one gets:
\begin{equation} \label{action heterotic}
S_{\text{Het}}=v \int d^{10} x \sqrt{g} e^{-2 \phi}\left[R+4 \partial_n \phi \partial^n \phi-\frac{1}{12} H_{m n k} H^{m n k}-\frac{1}{4} \delta_{\alpha \beta} F_{m n}{}^\alpha F^{m n \beta}\right] \, ,
\end{equation}
where $R$ is the 10-dimensional Ricci scalar. Equations of motion derived from this read: 
\begin{equation} \label{eoms heterotic}
    \begin{aligned}
        & R_{m n} - \dfrac{1}{4}H_{m k l} H_{n r s} g^{k r} g^{l s} + 2 \nabla_{m}{\nabla_{n}{\phi}} - \dfrac{1}{2}F_{m k \alpha} F_{n l}{}^{\alpha} g^{k l}  = 0 \, , \\
        & \nabla^{n}{(e^{-2\phi}H_{n m l})} = 0 \, , \\
        & \nabla^{n}{(e^{-2\phi} F_{m n \alpha})} + e^{-2\phi} f_{\alpha \beta \gamma} F_{m l}{}^{\beta} A^{l \gamma} - \dfrac{1}{2} e^{-2\phi}  F^{n l}{}_{\alpha} H_{m n l} = 0 \, , \\
        & R + 4 \nabla_{n}\nabla^{n}\phi - 4 \nabla_{n}\phi \nabla^{n}\phi - \dfrac{1}{12} H_{m n k}H^{m n k} - \dfrac{1}{4} F_{m n \alpha} F^{m n \alpha} = 0 \, .
    \end{aligned}
\end{equation}

It is useful to rewrite the action (\ref{action heterotic}) and the equations of motion (\ref{eoms heterotic}) in terms of algebra elements $A_{m} \equiv A_{m}{}^{\alpha} \mathfrak{t}_{\alpha} \,   $ , where $\mathfrak{t}_{\alpha}$ are the  generators of either SO(32) or E$_{8}\times$E$_{8}$ : 
\begin{equation}
S =v \int d^{10} x \sqrt{g} e^{-2 \phi}\left[R+4 \partial_n \phi \partial^n \phi-\frac{1}{12} H_{m n k} H^{m n k} + \frac{1}{4} \mathrm{Tr} (F_{m n} F^{m n})  \right] \, ,
\end{equation}
\begin{equation} 
    \begin{aligned}
        & R_{m n} - \dfrac{1}{4}H_{m k l} H_{n r s} g^{k r} g^{l s} + 2 \nabla_{m}{\nabla_{n}{\phi}} + \dfrac{1}{2} \mathrm{Tr} (F_{m k} F_{n l}) \, g^{k l}  = 0 \, , \\
        & \nabla^{n}{(e^{-2\phi}H_{n m l})} = 0 \, , \\
        & \nabla^{n}{(e^{-2\phi} F_{m n})} + e^{-2\phi}  [ F_{m l} , A^{l } ] - \dfrac{1}{2} e^{-2\phi}  F^{n l} H_{m n l} = 0 \, , \\
        & R + 4 \nabla_{n}\nabla^{n}\phi - 4 \nabla_{n}\phi \nabla^{n}\phi - \dfrac{1}{12} H_{m n k}H^{m n k} + \dfrac{1}{4} \mathrm{Tr} (F_{m n } F^{m n})  = 0 \, , 
    \end{aligned}
\end{equation}
with 
\begin{equation} 
    \begin{aligned}
        H_{m n k} =&\ 3 \partial_{[m} B_{n k]} + \dfrac{3}{2} \mathrm{Tr} (A_{[ m} F_{n k ]})    + \dfrac{1}{2} \mathrm{Tr} (A_{m} [A_{n} , A_{k} ] )  \, ,  \\
        F_{m n \alpha} =&\ 2 \partial_{[m} A_{n]} - [ A_{m} , A_{n} ] \, . 
    \end{aligned}
\end{equation}
Note that if the generators $\mathfrak{t}_{\alpha}$ are chosen such that $\mathrm{Tr}(\mathfrak{t}_{\alpha} \mathfrak{t}_{\beta}) = - \delta_{\alpha \beta} \, $ and $ [\mathfrak{t}_{\alpha} , \mathfrak{t}_{\beta} ] = f_{\alpha \beta \gamma} \mathfrak{t}_{\gamma} \, ,$ then the equations (\ref{eoms heterotic}) are reproduced.

\section{Bi- and uni-vector deformations}\label{sec:deformation}

To construct a deformation we follow the ideas of \cite{Bakhmatov:2019dow,Bakhmatov:2020kul,Gubarev:2020ydf,Barakin:2024rnz,Gubarev:2024tks,Gubarev:2025hvr}. We define a deformation as a simultaneous rotation of the generalized vielbein/metric and the structure constants
  \begin{equation} \label{Deformation}
      E'^A{}_{M} = O_{M}{}^{N}E^{A}{}_{N} \, , \quad \mathcal{H}'_{M N} = O_{M}{}^{R}O_{N}{}^{S}\mathcal{H}_{R S} \, ,  \quad f'_{MNK} = O_{M}{}^{P} O_{N}{}^{Q} O_{K}{}^{R} f_{P Q R}
  \end{equation}
with the O(D,D+n) matrix $O_{M}{}^{N}$ defined through the corresponding generators $T_{mn}$, $T_{m \alpha}$ of o(D,D+n)
\begin{equation} \label{deformation_matrix}
     O_{M}{}^{N} \, = \exp(\Sigma^{mn} T_{mn} + \gamma^{m \alpha} T_{m \alpha}) = \left(
   \begin{array}{lll} 
  \delta^{n}{}_m & 0 & 0\\
 \Sigma^{m n} & \delta_{n}{}^m &  \gamma^{m \beta} \\
 -\gamma^{n}{}_{\alpha} & 0 & \delta_{\alpha}{}^{\beta}
   \end{array}
   \right) 
   =  \left(
  \begin{array}{ccc} 
  \delta^{n}{}_m & 0 & 0\\
  \beta^{m n} - \dfrac{1}{2}\gamma^m{}_\a \gamma^{ n \a} & \delta_{n}{}^m & \gamma^{m \beta} \\
  -\gamma^{n}{}_{\alpha} & 0 & \delta_{\alpha}{}^{\beta}
   \end{array}
   \right), 
\end{equation}
where $\beta^{m n}$ is antisymmetric and $\Sigma^{m n} = \beta^{m n} - \dfrac{1}{2}\gamma^m{}_\a \gamma^{ n \a}$. This gives the following deformed generalized vielbein
\begin{equation}
    E'^A{}_{M} = \left(\begin{array}{ccc}
 e^a{}_m & -e_a{}^l c_{lm} & A_m{}^{\bar{\beta}} \\ 
 e^a{}_n \Sigma^{mn} & e_a{}^m - e_a{}^l A_{l \alpha } \gamma^{m \alpha} - e_a{}^l c_{ln} \Sigma^{mn} & A_n{}^{\bar{\beta}}\Sigma^{mn} + \delta_\beta{}^{\bar{\beta}}\gamma^{m \beta} \\
 -e^a{}_n \gamma^n{}_\alpha & - e_a{}^l A_{l \alpha} + e_a{}^l c_{ln} \gamma^n{}_{\alpha} & \delta_{\alpha}{}^{\bar{\beta}} - A_n{}^{\bar{\beta}} \gamma^n{}_{\alpha} 
    \end{array}\right) \, ,
\end{equation}
and generalized metric
\begin{equation}
    \mathcal{H}'_{M N} = E'^A{}_M E'^B{}_N S_{AB} \, .
\end{equation}
Comparing components of the latter with components of the generalized metric in the supergravity frame
\begin{equation} \label{generalized_metric_deformed}
     {\mathcal{H}}'_{M N} =
    \left(\begin{array}{ccc}
  \tilde{g}_{m n}+{\tilde{A}}_m{}^\gamma  {\tilde{A}}_{n \gamma }+{\tilde{c}}_{l m} {\tilde{g}}^{l s}  {\tilde{c}}_{s n}  &- {\tilde{g}}^{n l}  {\tilde{c}}_{l m} &  {\tilde{A}}_m{}_\beta + {\tilde{A}}_{l \beta }  {\tilde{g}}^{l s}  {\tilde{c}}_{s m} \\
 - {\tilde{g}}^{m l}  {\tilde{c}}_{l n}    &  {\tilde{g}}^{m n} & - {\tilde{g}}^{m l}  {\tilde{A}}_{ l \beta}  \\  {\tilde{A}}_{n \alpha} + {\tilde{A}}_{ l \alpha}  {\tilde{g}}^{l s}  {\tilde{c}}_{s n} & -  {\tilde{g}}^{n l}  {\tilde{A}}_{ l \alpha} & \delta_{\alpha \beta} + {\tilde{A}}_{ l \alpha}  {\tilde{g}}^{l s}  {\tilde{A}}_{ s \beta}
    \end{array}\right) \, ,
\end{equation}
we obtain the following deformation rules: 
\begin{equation} \label{deformation rules full}
    \begin{aligned}
          \mathcal{H}'_{mn} =  &\  \tilde{g}_{m n}+\tilde{A}_m{}^\gamma\tilde{A}_{n \gamma} + \tilde{c}_{l m } \tilde{g}^{l s} \tilde{c}_{s n} = g_{m n} + A_m{}^\gamma A_n{}_\gamma + c_{l m}g^{ls}c_{sn} \, ,\\
         \mathcal{H}'^m{}_n = &  -  \tilde{g}^{ml}\tilde{c}_{ln} = -g^{ml}c_{ln} + \Sigma^{mr} (g_{rn} + A_r{}^\gamma A_n{}_\gamma + c_{lr}g^{ls}c_{sn}) + \gamma^{m \beta} (A_n{}_\beta +A_{l \beta} g^{ls} c_{sn}) \, , \\
         \mathcal{H}'^{mn} = &\ \tilde{g}^{mn} = g^{mn} + \Sigma^{mr} \Sigma^{nk} (g_{rk} + A_r{}^{\gamma}A_k{}_\gamma +c_{lr}g^{ls}c_{sk} )  + \gamma^{m \alpha} \gamma^{n \beta} (\delta_{\alpha \beta} + A_{l \alpha} g^{ls}A_{s\beta}) \\
        & - 2 g^{l (m} \gamma^{n) \beta}   A_{l\beta} - 2 \Sigma^{(m |s}g^{|n)l} c_{ls}  + 2 \Sigma^{(m|r} \gamma^{|n) \beta} (A_{r\beta} + A_{l\beta} g^{ls}c_{sr})   \, ,\\
        \mathcal{H}'_{n \alpha} = &\ \tilde{A}_{n\alpha} + \tilde{A}_{l\alpha} \tilde{g}^{ls}\tilde{c}_{sn}  = A_{n\alpha} + A_{l \alpha} g^{ls} c_{sn} - \gamma^{r}{}_\alpha (g_{rn} + A_{r}{}^\gamma A_{n}{}_\gamma +c_{lr} g^{ls} c_{sn}) \, , \\
        \mathcal{H}'^n{}_\alpha = &\ -  \tilde{g}^{nl} \tilde{A}_{l\alpha} = - g^{nl}A_{l\alpha} - \Sigma^{nr} \gamma^{k}{}_\alpha (g_{rk}+A_r{}^\gamma A_k{}_\gamma +c_{lr}g^{ls}c_{sk})  + \gamma^k{}_\alpha  g^{nl}c_{lk}  \\ 
        & + \Sigma^{n r}(A_{r\alpha} + A_{l\alpha}g^{ls}c_{sr}) + \gamma^{n\beta} (\delta_{\alpha \beta} + A_{l \alpha } g^{ls} A_{s \beta}) \, , \\
        \mathcal{H}'_{\alpha \beta} = &\ \delta_{\alpha\beta} + \tilde{A}_{l\alpha}\tilde{g}^{ls}\tilde{A}_{s\beta} = \delta_{\alpha\beta} + A_{l\alpha}g^{ls}A_{s\beta} - \gamma^r{}_{\alpha} (A_{r \beta} + A_{l \beta}g^{ls}c_{sr}) \\
        & \qquad\qquad\qquad - \gamma^r{}_{\beta} (A_{r \alpha} + A_{l \alpha}g^{ls}c_{sr}) + \gamma^{r}{}_{\alpha}  \gamma^k{}_\beta (g_{rk} + A_r{}^\gamma A_{k \gamma} + c_{lr} g^{ls}c_{sk} )    \, ,
    \end{aligned}
\end{equation}
Therefore the recipe for the construction of a deformation is the following: 1) recover the deformed inverse metric $\tilde{g}^{mn}$ from $\mathcal{H}'^{mn}$, 2) recover the $\tilde{c}_{ln}$ from $\mathcal{H}'^m{}_n$, 3) recover the $\tilde{A}_{l\alpha}$ from $\mathcal{H}'^n{}_\alpha$, 4) recover the $\tilde{b}_{mn}$ from $\tilde{c}_{ln}$ and the known $\tilde{A}_{l\alpha}$. The remained equations $\mathcal{H}'_{mn},\mathcal{H}'_{n \alpha},\mathcal{H}'_{\alpha \beta}$ are satisfied automatically as $\mathcal{H}'$ is in O(D,D+n). As a result we obtain the following deformation rules:
\begin{equation}  \label{deformed fields}
    \begin{aligned}
       \ \tilde{g}^{mn} =&\ g^{mn} + \Sigma^{mr} \Sigma^{nk} (g_{rk} + A_r{}^{\gamma}A_k{}_\gamma +c_{lr}g^{ls}c_{sk} )  + \gamma^{m \alpha} \gamma^{n \beta} (\delta_{\alpha \beta} + A_{l \alpha} g^{ls}A_{s\beta}) \\
        &  -  2 g^{l(n} \gamma^{m) \alpha} A_{l\alpha} - 2 g^{l(m} \Sigma^{n)s}c_{ls} + 2 \Sigma^{(m|r} \gamma^{|n)\beta} (A_{r\beta} + A_{l\beta} g^{ls}c_{sr}) \, ,\\
            \tilde{c}_{m n} =&\ g^{ml}c_{ln} - \tilde{g}_{m k} \Sigma^{k r} (g_{rn} + A_r{}^\gamma A_n{}_\gamma + c_{lr}g^{ls}c_{sn}) - \tilde{g}_{m k} \gamma^{k \beta} (A_n{}_\beta +A_{l \beta} g^{ls} c_{sn}) \, , \\         
            \tilde{A}_{m \alpha} =&\   \tilde{g}_{m n}  g^{n l}A_{l\alpha} +  \tilde{g}_{m n}  \Sigma^{nr} \gamma^{k}{}_\alpha (g_{rk}+A_r{}^\gamma A_k{}_\gamma +c_{lr}g^{ls}c_{sk})  -  \tilde{g}_{m n}  \gamma^k{}_\alpha  g^{nl}c_{lk} \\ 
        &  -  \tilde{g}_{m n} \Sigma^{n r}(A_{r\alpha} + A_{l\alpha}g^{ls}c_{sr}) -  \tilde{g}_{m n}  \gamma^{n\beta} (\delta_{\alpha \beta} + A_{l \alpha } g^{ls} A_{s \beta}) \, , 
    \end{aligned}
\end{equation}
or, in terms of the Lie algebra elements:
\begin{equation}  \label{deformed fields}
    \begin{aligned}
       \ \tilde{g}^{mn} =&\ g^{mn} + \Sigma^{mr} \Sigma^{nk} (g_{rk} - \mathrm{Tr}(A_r A_k)  + c_{lr}g^{ls}c_{sk} )   - \mathrm{Tr} \, (\gamma^{m} \gamma^{n})   + g^{ls} \,\mathrm{Tr} \, (\gamma^{m} A_{l})  \; \mathrm{Tr} \,(\gamma^{n} A_{s})    \\
        &   +   2 g^{l (m} \, \mathrm{Tr} \, (\gamma^{n)} A_{l})  - 2 \Sigma^{(m|s}g^{|n) l} c_{ls} - 2 \Sigma^{(m| r}  ( \mathrm{Tr} \, (\gamma^{|n)} A_{r}) + g^{ls}c_{sr} \, \mathrm{Tr} \, (\gamma^{|n)} A_{l}) )   \, ,\\
        \tilde{c}_{m n} =&\ g^{ml}c_{ln} - \tilde{g}_{m k} \Sigma^{k r} (g_{rn} - \mathrm{Tr} (A_{r} A_{n}) + c_{lr}g^{ls}c_{sn}) + \tilde{g}_{m k} ( \mathrm{Tr} \, (\gamma^{k} A_{n}) + g^{l s} c_{s n} \, \mathrm{Tr} \, (\gamma^{k} A_{l}) ) \, , \\         
        \tilde{A}_{m} =&\   \tilde{g}_{m n}  g^{n l}A_{l} +  \tilde{g}_{m n}  \Sigma^{nr} \gamma^{k} (g_{rk} - \mathrm{Tr} (A_{r} A_{k})  + c_{lr}g^{ls}c_{sk})  -  \tilde{g}_{m n}  \gamma^{k}  g^{nl}c_{lk}   \\ 
        & -  \tilde{g}_{m n} \Sigma^{n r}(A_{r} + A_{l}g^{ls}c_{sr}) -  \tilde{g}_{m n}  \gamma^{n}  +  \tilde{g}_{m n}    A_{l} g^{ls} \, \mathrm{Tr} \, (\gamma^{n} A_{s}) \, , 
    \end{aligned}
\end{equation}
with $\Sigma^{m n} = \beta^{m n} + \dfrac{1}{2} \mathrm{Tr} \, (\gamma^{m} \gamma^{n}) \, $. 

It is worth to mention a subtlety concerning the definition of the deformed generalized metric (\ref{generalized_metric_deformed}). It's most general form is given by
\begin{equation}
     {\mathcal{H}}'_{M N} =
    \left(\begin{array}{ccc}
  \tilde{g}_{m n}+{\tilde{A}}_m{}^\alpha M_{\alpha \beta}  {\tilde{A}}_{n}{}^{\beta }+{\tilde{c}}_{l m} {\tilde{g}}^{l s}  {\tilde{c}}_{s n}  &- {\tilde{g}}^{n l}  {\tilde{c}}_{l m} & M_{\beta \gamma} {\tilde{A}}_m{}^{\gamma} + {\tilde{A}}_{l \beta }  {\tilde{g}}^{l s}  {\tilde{c}}_{s m} \\
 - {\tilde{g}}^{m l}  {\tilde{c}}_{l n}    &  {\tilde{g}}^{m n} & - {\tilde{g}}^{m l}  {\tilde{A}}_{ l \beta}  \\ 
M_{\alpha \gamma} {\tilde{A}}_{n}{}^{\gamma} + {\tilde{A}}_{ l \alpha}  {\tilde{g}}^{l s}  {\tilde{c}}_{s n} & -  {\tilde{g}}^{n l}  {\tilde{A}}_{ l \alpha} & M_{\alpha \beta} + {\tilde{A}}_{ l \alpha}  {\tilde{g}}^{l s}  {\tilde{A}}_{ s \beta}
    \end{array}\right) \, ,
\end{equation}
with a symmetric matrix $M_{\alpha \beta}$. The condition $\mathcal{H}' \in$ O(D,D+n) requires that $M_{\alpha \gamma} M_{\beta \gamma}  = \delta_{\alpha \beta}$, i.e. $M$ is an orthogonal matrix. On the other hand, $M_{\alpha \beta}$ being both symmetric and orthogonal means that its infinitesimal transformation $\delta M_{\alpha \beta}$ such that,  $M_{\alpha \beta} = \delta_{\alpha \beta} + \delta M_{\alpha \beta} \, ,$ is both symmetric and antisymmetric in the indices $\alpha$ and $\beta$, i.e. it is zero. Therefore, we argue that $M_{\alpha \beta} = \delta_{\alpha \beta}$ for any continuous deformation, including (\ref{deformation_matrix}).

Let us now comment on the transformation of $f_{M N K}$ in (\ref{Deformation}). The only non-zero components of $f^{M N K}$ are $f^{\alpha \beta \gamma} \, $, which are the fixed constants in our theory. However, their transformation yields additional components  $f'^{m \beta \gamma } = \gamma^{m}{}_{\alpha} f^{\alpha \beta \gamma}$, $ \; f'^{m n \gamma } = \gamma^{m}{}_{\alpha} \gamma^{n}{}_{\beta} f^{\alpha \beta \gamma}$, $ \; f'^{m n k } = \gamma^{m}{}_{\alpha} \gamma^{n}{}_{\beta} \gamma^{k}{}_{\gamma} f^{\alpha \beta \gamma}$. To remain in the heterotic theory the contribution of these components should vanish in the equations of motion.  Such terms enter the equations of motion and the action only as commutators $[\gamma^{m}, A_{n}]$ and $[\gamma^{m}, \gamma^{n}]$, where $\gamma^{m} = \gamma^{m}{}_{\alpha} \mathfrak{t}^{\alpha}$ and $A_{m} = A_{m \alpha} \mathfrak{t}^{\alpha}$. Hence for the deformation (\ref{deformation_matrix}) to work, we need $\gamma^{m}{}_{\alpha}$ to satisfy
\begin{equation} \label{commuting_conditions}
    [\gamma^{m}, A_{n}] = 0 \, , \quad [\gamma^{m}, \gamma^{n}] = 0 \, .
\end{equation}
The transformation for the dilaton can be found from the transformation of the invariant dilaton that is
\begin{equation}
    d = \sqrt{g} e^{-2\phi} = \sqrt{\tilde{g}} e^{-2\tilde{\phi}} = d' .
\end{equation}
Let us now rewrite equations (\ref{deformation rules full}) in matrix form for the case when $A_{m\alpha} = b_{mn} = 0$ :
\begin{equation}
\begin{aligned}\label{eq1}
    \tilde{g} + \tilde{A}\tilde{A}^T + \tilde{c}^T\tilde{g}^{-1}\tilde{c} =&\ g \, , \\
    - \tilde{g}^{-1} \tilde{c} =&\ \Sigma g\, , \\
    \tilde{g}^{-1} =&\  g^{-1} + \Sigma g \Sigma^T + \gamma \gamma^T \, , \\
    - \tilde{g}^{-1} A =&\ - \Sigma g \gamma + \gamma \, , \\
    \tilde{A} + \tilde{c}^T \tilde{g}^{-1}A =&\ - g \gamma \, , \\
    \tilde{A}^T \tilde{g}^{-1} A =&\ \gamma^T g \gamma \, .
\end{aligned}
\end{equation}
Analogous to the case of $\beta$-deformation from the first two equations we get a generalization of the open/closed map: 
\begin{equation} \label{open/closed}
    ( \tilde{g} + \tilde{c}^T ) (g^{-1} - \Sigma) = 1 \, , 
\end{equation}
The derivation is straightforward: 
\begin{equation}
    \begin{aligned}
      &  (\tilde{g} + \tilde{c}^T)(g^{-1} - \Sigma) = (g - \tilde{A}\tilde{A}^T - \tilde{c}^T\tilde{g}^{-1}\tilde{c})g^{-1} - \tilde{c}^T\Sigma + \tilde{c}^T g^{-1} - \tilde{g}\Sigma \,  =  \\
      & = 1 - \tilde{A}\tilde{A}^Tg^{-1} - \tilde{c}^T\tilde{g}^{-1}\tilde{c}g^{-1} - \tilde{c}^T (-\tilde{g}^{-1}\tilde{c}g^{-1} ) + \tilde{c}^Tg^{-1} + \tilde{c}g^{-1} = \\
      & = 1 - \tilde{A}\tilde{A}^T g^{-1} + (\tilde{c}^T+ \tilde{c}) g^{-1} =1 \, ,
    \end{aligned}
\end{equation}
where we have used that $\Sigma = -  \tilde{g}^{-1}\tilde{c}g^{-1}$ and $\tilde{c}^T + \tilde{c} = \tilde{A}\tilde{A}^T$ . In the case of GDFT the open/closed map determines only $\tilde{b}$ and $\tilde{g} + \dfrac{1}{2} \tilde{A} \tilde{A}^T \, . $ To determine $\tilde{A}$ we can take the last three equations of (\ref{eq1}):  
\begin{equation}
    \begin{aligned}
    &    \tilde{A} +\tilde{c}^T\tilde{g}^{-1}\tilde{A} = \tilde{A} - \tilde{b}\tilde{g}^{-1}\tilde{A} + \dfrac{1}{2} \tilde{A} \tilde{A}^{T} \tilde{g}^{-1} \tilde{A} = \tilde{A} + \tilde{b}(-\Sigma g \gamma + \gamma) + \frac{1}{2} \tilde{A} \gamma^T g \gamma = - g \gamma \\
    \end{aligned}
\end{equation}
\begin{equation} \label{A_transform}
     \Rightarrow \, \tilde{A}(1+\dfrac{1}{2}\gamma^Tg\gamma) = \tilde{b}(\Sigma g \gamma - \gamma) - g\gamma \, ,  
\end{equation}
which determines $\tilde{A}$ provided that $\tilde{b}$ is already determined. Finally, for the deformation parameters we use poly-Killing ansatz
\begin{equation}
    \beta^{mn} = r^{ij} k_{i}{}^{m} k_{j}{}^{n}, \quad \gamma^m = \rho^{i}{}_{\alpha} k_{i}{}^{m} \mathfrak{t}^{\alpha}.
\end{equation}
In what follows we assume  that the conditions (\ref{10dcond}) are sufficient for the deformation to generate solutions, however we leave the rigorous proof of that for the future work. Instead here we will focus at the explicit examples of such deformations.

\section{Examples}\label{sec:examples}

Now we will consider examples of the above deformations and take an SO(2) subgroup of either SO(32) or E$_{8}\times$E$_{8}$ and in what follows use the generator
\begin{equation}\label{so2generator}
    \mathfrak{t} = \mathfrak{t}_{3} = \frac{i \sigma_{3}}{\sqrt{2}} = \frac{i}{\sqrt{2}}
    \begin{pmatrix}
        1 & 0\\
        0 & -1
    \end{pmatrix}.
\end{equation}

\subsection{Flat space uni-vector deformation}
Let us first consider a deformation of the flat background:  
 \begin{equation} \label{flat_space}
      ds^2 = -dt^2 + dr^2 + r^2 d\theta^2 + \sum_{i=4}^{10} dx^i dx^i \, ,  \quad \phi = 0 \, , \quad b_{m n} =0 \, , \quad A_{m \alpha} = 0  \, .
 \end{equation}
We take $\gamma = \sqrt{2} \, \eta  \, \mathfrak{t} \, \partial_{\theta} \, , $  where $\eta$ is a parameter of the deformation and $\mathfrak{t} $ is generator (\ref{so2generator}) in the SO(32) or $E_8 \times E_8$ algebra, such that $\mathrm{Tr} \, \mathfrak{t}^2 = -1 \, .$ Note that, being proportional to a U(1) generator $\gamma$ satisfies (\ref{commuting_conditions}).

The deformed fields read 
\begin{equation}
    \begin{aligned}
    ds^2 =&\ -dt^2 + dr^2 + \dfrac{r^2}{(1+\eta^2 r^2)^2} d\theta^2  + \sum_{i=4}^{10} dx^i dx^i \, , \\ 
   \tilde{\phi} =&\ - \dfrac{1}{2} \mathrm{ln}(1+\eta^2 r^2) \, , \\
   \tilde{b}_{m n} =&\ 0 \, , \\
   \tilde{A} =&\ - \eta \, \dfrac{\sqrt{2} \, r^2}{1+\eta^2 r^2} \, \mathfrak{t} \, d\theta  \, .
    \end{aligned}
\end{equation}
Such transformation is nontrivial, because the Ricci scalar and the field strength for $A$ become non-zero and are respectively
\begin{equation}
    \tilde{R} = - \dfrac{4\eta^2(-3+\eta^2r^2)}{(1+\eta^2r^2)^2} \, , \quad  \tilde{F} = - \dfrac{2 \sqrt{2} \, \eta \, r}{(1+\eta^2 r^2)^2} \, \mathfrak{t} \, dr \wedge d\theta \, .
\end{equation}

\subsection{Flat space uni- and bi-vector deformation}
Consider now the deformation of the flat space (\ref{flat_space}) with poly-vectors $\gamma = \sqrt{2} \, \eta \, \mathfrak{t} \, \partial_\theta$ and $\beta =  \, \omega \, \partial_\theta \wedge \partial_t$, where $\eta$ and $\omega$ are parameters of the deformation. The deformed fields take the following form 
\begin{equation}
    \begin{aligned}
    ds^2 =&\ \dfrac{-1 + (-2\eta^2 + \omega^2)\, r^2 - \eta^4 \, r^4 }{(U(r))^2} dt^2 + dr^2 + \dfrac{r^2 - \omega^2 r^4}{(U(r))^2} d\theta^2 + \dfrac{2 \eta^2 \omega \, r^4}{(U(r))^2} dt d\theta  + \sum_{i-4}^{10} dx^i dx^i \, , \\
    \tilde{\phi} =&\ - \dfrac{1}{2} \mathrm{ln}(U(r)) \, , \\
    \tilde{b} =&\ \dfrac{ \omega \, r^2}{U(r)}  d\theta  \wedge dt \, , \\
    \tilde{A} =&\  \dfrac{\eta \, \sqrt{2} \, r^2}{U(r)} \mathfrak{t} \, dt  -  \dfrac{\eta \, \sqrt{2}  \, r^2}{U(r)} \mathfrak{t} \, d\theta \, ,\\
    U(r) =&\ 1+(\eta^2-\omega^2) \, r^2 .
    \end{aligned}
\end{equation}
 The Ricci scalar and the field strengths are again non-trivial and have the following form:
\begin{equation}
    \begin{aligned}
       \tilde{R} =&\ \frac{12 \eta ^2-4 r^2 \left(\eta ^2-\omega ^2\right)^2-10 \omega ^2}{(U(r))^2} \, ,  \\
       \tilde{F} =&\ - \dfrac{2 \sqrt{2}  \, \eta \, r}{(U(r) )^2} \mathfrak{t} \, dr \wedge d\theta + \dfrac{2 \sqrt{2}  \, \eta  \, \omega \, r}{(U(r) )^2} \mathfrak{t} \, d\theta \wedge dt \, , \\
       \tilde{H} =&\ \dfrac{2 \omega \, r }{U(r))^2} \, dt \wedge dr \wedge d\theta \, . 
    \end{aligned}
\end{equation}

\subsection{F1-string uni-vector deformation}
Next consider the F1-string solution:
\begin{equation} \label{F1_string}
    \begin{aligned}
     &   ds^2 = H(r)^{-1}(- dt^2 + dx^2) + dr^2 + r^2ds^2_{\mathbb{S}^7} \, , \\
     & e^{-2\phi} = H(r) \, , \\
     &   b = - (H(r)^{-1} - 1) \, dt \wedge dx\, , \\
     & A = 0 \, , 
    \end{aligned}
\end{equation}
where $H(r) = 1 + \dfrac{R^6}{r^6} \, .$ We deform this solution by the vector   $ \gamma = \sqrt{2} \, \eta \, \mathfrak{t} \, \partial_t \, ,$ where $\eta$ is the deformation parameter and $\mathfrak{t}$ is given by (\ref{so2generator}). The resulting deformed solution is
\begin{equation}
    \begin{aligned}
        ds^2 =&\ - \frac{ R^6 \,r^6+r^{12}}{(U(r))^2} dt^2 + 2\dfrac{\eta^2 \, r^6 \, R^6}{(U(r))^2} dt \, dx \\
        & \qquad \qquad + \dfrac{r^{12}(\eta^2-1)^2-r^6 R^6(\eta^4-1)}{(U(r))^2}dx^2 + dr^2 + r^2ds^2_{\mathbb{S}^7} \, , \\
        \tilde{\phi} =&\ \dfrac{1}{2} \mathrm{ln}\dfrac{r^{6}}{U(r)}  \, , \\
        \tilde{A} =&\ - \dfrac{\sqrt{2} \, \eta \, r^6}{U(r)} \mathfrak{t} \, dt - \dfrac{\sqrt{2} \, \eta \, R^6}{U(r)} \mathfrak{t} \, dx \, , \\ 
        \tilde{b} =&\ \dfrac{R^6}{U(r)}  \, dt \wedge dx \, , \\
        U(r) =&\ R^6 + r^6(1-\eta^2)  \, ,
    \end{aligned}
\end{equation}
and expressions for the non-trivial field strengths can be found in the Wolfram Mathematica files of \cite{Gubarev:2025heterotic}.

\subsection{Singular F1-string uni- and bi-vector deformation}
We can also perform both bi- and uni-vector deformations of the F1 background (\ref{F1_string}) simultaneously. A bi-vector deformation $\beta =  \, \omega \, \partial_t \wedge \partial_x $ is known to be singular for $\omega = -1 \, $ \cite{Seiberg:2000ms}. However, together with uni-vector deformation $ \gamma = \sqrt{2} \, \eta \, \mathfrak{t} \, \partial_t \, ,$ it becomes regular for $\eta \neq \, 0$. The deformed solution for $\omega = -1$ reads:
\begin{equation}
    \begin{aligned}
    ds^2 =&\ \dfrac{2}{\eta^2} dt dx + \left(1- \dfrac{2}{\eta^2} - \dfrac{R^6}{r^6}\right) dx^2 + dr^2 + r^2 ds_{\mathbb{S}^7} \, , \\
   \tilde{\phi} =&\ - \mathrm{ln} \, \eta \, , \\
   \tilde{b} =&\ - \dfrac{1}{\eta^2} dt \wedge dx \, , \\
   \tilde{A} =&\ \dfrac{\sqrt{2}}{\eta} \, \mathfrak{t} \,  dt + \dfrac{\sqrt{2}}{\eta} \, \mathfrak{t} \, dx \, .
    \end{aligned}
\end{equation}
After gauging away the constant fields we are left with a solution: 
\begin{equation}
    \begin{aligned}
    ds^2 =&\ \dfrac{2}{\eta^2} dt dx + \left(1- \dfrac{2}{\eta^2} - \dfrac{R^6}{r^6}\right) dx^2 + dr^2 + r^2 ds_{\mathbb{S}^7} \, , \\
   \tilde{\phi} =&\ 0 \, , \\
   \tilde{b} =&\ 0 \, , \\
   \tilde{A} =&\ 0  \, , 
    \end{aligned}
\end{equation}
which is Ricci-flat, but has a non-vanishing Riemann tensor, see \cite{Gubarev:2025heterotic}.

\subsection{F1-string uni- and bi-vector deformation}
Now we perform  bi- and uni-vector deformations of F1 background (\ref{F1_string}) simultaneously with a non-singular $\beta = \omega \, \partial_t \wedge \partial_x $ and $ \gamma = \sqrt{2} \, \eta \, \mathfrak{t} \, \partial_t \, $. The deformed solution reads:
\begin{equation}
    \begin{aligned}
        ds^2 =&\ - \frac{ R^6 \,r^6(1+\omega)^2 +r^{12}(1-\omega^2)}{(U(r))^2} dt^2 + 2\dfrac{\eta^2 \, r^6 \, R^6(1+\omega^2) - \eta^2 \omega \, r^{12}}{(U(r))^2} dt \, dx \\
        & \quad + \dfrac{r^{12}((\eta^2-1)^2-\omega^2)-r^6 R^6(\eta^4-\omega- 1)}{(U(r))^2}dx^2 + dr^2 + r^2ds^2_{\mathbb{S}^7} \, , \\
        \tilde{\phi} =&\   \dfrac{1}{2} \mathrm{ln}\dfrac{r^{6}}{U(r)}  \, , \\
        \tilde{A} =&\ - \dfrac{\sqrt{2} \, \eta \, r^6}{U(r)} \mathfrak{t} \, dt - \dfrac{\sqrt{2} \, \eta \, ( R^6 (1+\omega) - r^6 \omega)}{U(r)} \mathfrak{t} \, dx \, , \\ 
        \tilde{b} =&\ \dfrac{R^6 (1+\omega) - r^6 \, \omega}{U(r)}  \, dt \wedge dx \, , \\
        U(r) =&\ R^6 (1+\omega)^2 + r^6\left(1-\eta^2-\omega^2\right) \, .
    \end{aligned}
\end{equation}
which also has a non-vanishing Riemann tensor and non-trivial fluxes, see \cite{Gubarev:2025heterotic}.

\section{Conclusion and discussion}\label{sec:conclusion}

In this work, we have constructed uni- and bi-vector deformations of heterotic supergravity solutions. To achieve this, we extended the approach of building poly-vector deformations via extended field theories to the case of gauged double field theory (GDFT). The deformations were defined as a local O(D,D+n) rotations (\ref{Deformation}), (\ref{deformation_matrix}) with the deformation parameters taken to be poly-Killing. We derived the explicit deformation rules (\ref{deformed fields}), which generalize the open/closed string map. Finally, we constructed explicit examples of uni- and bi-vector deformations for the flat space and the fundamental string (F1) solutions.

As constraints ensuring that the deformations generate supergravity solutions, we derived the condition that the uni-vector parameters must be commuting as algebra elements:
\begin{equation} \label{commuting_conditions_concl}
    [\gamma^{m}, A_{n}] = 0 \, , \quad [\gamma^{m}, \gamma^{n}] = 0 \, .
\end{equation}
We also assume that the $r$-matrix in $\beta^{mn} = r^{ij} k_{i}{}^{m} k_{j}{}^{n}$ satisfies
\begin{equation}\label{10dcond_concl}
    \begin{cases}
         f_{j_1 j_2}\,^{[i_1} r^{i_2|j_1|} r^{i_3]j_2} = 0, \qquad &\text{(CYBE)},\\
         f_{i_1 i_2}\,^{j} r^{i_1 i_2} = 0, \qquad &\text{(unimodularity)},
    \end{cases}
\end{equation}
which holds for all examples considered above. However, a rigorous proof of sufficiency of the conditions (\ref{10dcond_concl}) for the deformed backgrounds to be solutions remains a task for future work.

The new class of deformations considered here open up several interesting questions for future investigation. The first is the question of the string-theoretic or geometric interpretation of the uni-vector deformation parameter. For bi- and tri-vector deformations in the supergravity case, it is known that constant deformation poly-vectors $\beta^{ij},\Omega^{ijk}$ are related respectively to the non-commutativity of string endpoints \cite{Seiberg:1999vs}
\begin{equation}\label{nonc1}
    \langle [X^i(\tau), X^j(\tau')] \rangle = {i\over 2} \beta^{ij}
\epsilon(\tau-\tau') \, ,
\end{equation}
and to a loop algebra for the closed ends (``loops'') of M2-branes \cite{Berman:2004jv}
\begin{equation}\label{nonc2}
[ \tilde X^i(\sigma), \tilde X^j(\sigma') ]
= \underbrace{\Theta \epsilon^{ijk}}_{\Omega^{ijk}} \partial_\sigma X_k \delta(\sigma-\sigma') \, .
\end{equation}
The relevant question is: does there exist a generalization of (\ref{nonc1}) and (\ref{nonc2}) for uni-vector deformations? A similar question is also interesting for uni-vector deformations in the Einstein--Maxwell-dilaton theory \cite{Gubarev:2025hvr}.

The second promising direction is the holographic application of the constructed deformations. While heterotic theory lacks D-branes, it does contain NS5-branes. The Callan--Harvey--Strominger (CHS) model \cite{Callan:1991at}, which describes the near-horizon geometry of NS5-branes, serves as an archetypal gravitational background. Holographic duality \cite{Aharony:1998ub,Fotopoulos:2007rm,Monnee:2025msf} states that string theory in this background is dual to the six-dimensional little string theory (LST) \cite{Seiberg:1997zk} — a highly non-local and T-duality invariant theory describing the decoupled dynamics on the brane's worldvolume. The deformations constructed in this work can be applied to the NS5-brane background and are expected to correspond holographically to adding (ir)relevant or marginal operators, introducing non-commutativity, or yielding a non-vanishing VEV for an operator \cite{Fotopoulos:2010wc}. This formalism could provide a new tool for exploring the landscape of LSTs.

The third interesting direction is to investigate the preservation of integrability and kappa-symmetry for the two-dimensional sigma model on the deformed heterotic backgrounds, which also has direct holographic implications for LSTs. For these purposes quite useful approach is gauged double sigma model \cite{Hatsuda:2022zpi}. Furthermore, the recent work \cite{Osten:2023cza} has explored integrable heterotic deformations of the principal chiral model that turn on gauge fields. A natural and intriguing question is whether this deformation can be interpreted as a uni-vector heterotic deformation. Also, recent work \cite{Musaev:2025zww} has taken the first steps toward understanding integrability for poly-vector deformed backgrounds.

The fourth intriguing direction of arises from the relation between constant bi-vector deformations and $T\bar{T}$-deformations \cite{Blair:2020ops}. A natural question is: what kind of deformation does a uni-vector deformation correspond to? This question remains equally interesting for non-constant uni- and bi-vectors and for higher poly-vector deformations, such as tri-, quadri-, and six-vectors.

Finally, the ``sedimentation'' effect for poly-vector deformations of supergravity solutions was recently observed in \cite{Barakin:2025jwp}, that in a few words states that turning on poly-vector deformation can be understood as ``sedimentation'' (addition) of some physical objects to the initial background. This raises a compelling question: the ``sedimentation'' of which physical objects in heterotic supergravity does the uni-vector deformation correspond to?

\section{Acknowledgments}

We are grateful to Edvard Musaev for his valuable comments on the text and discussions of related topics. We also thank Sergei Barakin for discussions on related topics. The work of Kirill Gubarev was supported by the state assignment of the Institute for Information Transmission Problems of RAS.

\newpage

\bibliographystyle{utphys}
\bibliography{bib}

\end{document}

%% file: Defs.tex
\def\a{\alpha} \def\b{\beta}

\def\ph{\phantom}

\def\mH{\mathcal{H}}

\newcommand\bqa {\begin{eqnarray}}
\newcommand\eqa {\end{eqnarray}}

\newcommand{\bear}{\begin{array}}
\newcommand{\enar}{\end{array}}

\newcommand{\be}{\begin{equation}}
\newcommand{\ee}{\end{equation}}
\newcommand{\bea}{\begin{eqnarray}}
\newcommand{\eea}{\end{eqnarray}}